\documentclass[twocolumn,aps,pra,showpacs,amsmath]{revtex4}
\usepackage{amssymb}
\usepackage{mathrsfs}
\usepackage{graphicx}
\usepackage{float}
\usepackage{txfonts}
\usepackage{bm}
\usepackage{color}

\makeatletter

\usepackage{ulem}
\makeatother

\begin{document}
\title{Strain induced topological phase transition at zigzag edges of monolayer transition-metal dichalcogenides}
\author{Linhu Li,$^{1,2}$ Eduardo V. Castro,$^{1,2}$ and Pedro D. Sacramento$^{1,2}$}
\affiliation{$^{1}$Beijing Computational Science Research Center, Beijing 100084,
China}
\affiliation{$^{2}$CeFEMA, Instituto Superior T\'{e}cnico, Universidade de Lisboa,
Av. Rovisco Pais, 1049-001 Lisboa, Portugal}

\begin{abstract}
The effect of strain in zigzag ribbons of monolayer transition-metal dichalcogenides with induced superconductivity is studied using a minimal 3-band tight-binding model. The unstrained system shows a topological phase with Majorana zero modes localized at the boundaries of the one-dimensional (1D) zigzag edges.  By direct inspection of the spectrum and wave functions we examine the evolution of the topological phase as an in-plane, uniaxial deformation is imposed. It is found that strain shifts the energy of 1D edge states, thus causing a topological phase transition which eliminates the Majorana modes. For realistic parameter values, we show that the effect of strain can be changed from completely destructive -- in which case a small built in strain is enough to destroy the topological phase -- to a situation where strain becomes an effective tuning parameter which can be used to manipulate Majorana zero modes. These two regimes are accessible by increasing the value of the applied Zeeman field within realistic values. We also study how strain effects are affected by the chemical potential, showing in particular how unwanted effects can be minimized. Finally, as a cross-check of the obtained results, we reveal the connection between 1D Majorana zero modes in the zigzag edge and the multi-band Berry phase, which serves as a topological invariant of this system. 
\end{abstract}

\pacs{}
\maketitle
\date{today}

\section{Introduction}

A Majorana mode is a state at zero energy which is the conjugate of itself. These modes satisfy non-Abelian exchange statistics, and are considered strong candidates for fault-tolerant topological quantum computation \cite{TQC}. It has been proposed that Majorana zero modes can be realized in one-dimensional (1D) nanowires with spin-orbit coupling (SOC) and magnetic field, placed on top of an $s$-wave superconductor \cite{Lutchyn,Oreg}, and the experimental realization came just a few years later \cite{SC_experiment,SC_experiment2,SC_experiment3,SC_experiment4}.

During the last few years, much interest has been drawn to the monolayers of two-dimensional transition-metal dichalcogenides (TMD) of $\mathrm{MX_2}$ such as $\mathrm{MoS_2}$ or $\mathrm{WSe_2}$ \cite{XXH14}. These materials are semiconductors with a band gap of about 1~eV, which makes them good candidates for electronic and optoelectronic applications \cite{TMD}. Recently, it has been shown by different groups \cite{TMD_spin_SC,TMD_Majorana} that the zigzag edges of monolayer TMDs provide a promising platform for generating 1D Majorana modes. The graphene-like honeycomb lattice structures of these materials support several single band edge states, which are essential for generating 1D Majorana modes, and they also present a strong spin-orbit coupling (SOC), which is important for robust 1D topological superconductors \cite{Potter,Sau}. As a comparison to graphene, the sizable band gap of TMDs make it easier to study the 1D edge modes within the gap.

Beyond these advantages, TMDs also have an outstanding stretchability, and their physical properties can be greatly changed by strain. For example, a $2\% -3\%$ uniaxial/biaxial tensile strain can cause a direct-to-indirect band gap transition \cite{Feng,Wang}. A band gap reduction through strain has been demonstrated experimentally, and the consequent funneling of excitons detected \cite{Roldan13}. Exciton confinement through strain gradients has also been measured recently \cite{Kumar15}. Moreover, larger tensile biaxial strain of $10\% -15\%$ can drive the system into metallic phase \cite{Scalise,G-Asl}. It has also been demonstrated that strained TMDs may support a two-dimensional time-reversal-invariant topological phase \cite{Ochoa}. Although more realistic models have shown that such phase may be absent \cite{RRC+15}, they have also shown that spin manipulation through strain is possible due to the spin strain coupling.

Given the demonstrated sensitivity to strain in monolayer TMDs, an interesting question arising here is how strain affects the 1D edge states and the corresponding Majorana modes. The question is even of practical importance, since under realistic experimental conditions there may be small uncontrollable strain which will affect the system. In graphene, built in strain of order $0.01\% - 0.1\%$ has been reported in suspended samples \cite{COK+10,OCK+12}. In TMDs, residual strain $\lesssim \pm 0.05\%$ cannot be ruled out in current experiments \cite{Roldan13,Kumar15,RYZ+13}, in particular for monolayer samples. Here we show that by controlling the applied in-plane magnetic (Zeeman) field and the chemical potential, the effect of strain can be divided in three different regimes: (i) completely destructive for small magnetic fields, with Majorana physics totally washed out; (ii) strain engineering for moderate magnetic fields, where Majorana physics can be turned on and off through strain; (iii) strain  made irrelevant if the chemical potential is conveniently tuned.

The paper is organized as follows: In Sec.~\ref{Model}, we introduce the minimum model used to describe TMDs in the Majorana physics regime, and explain how strain is incorporated into the model. In Sec.~\ref{Berry}, we calculate the Majorana zero modes induced by applying superconductivity and an in-plane magnetic field, and define the multi-band Berry phase as the topological invariant to characterize the system. The effect of strain on the topological phase, and how it competes with the applied magnetic field and also with changes of the chemical potential is presented in Sec.~\ref{Results}. Concluding remarks are given in Sec.~\ref{Conclusions}.

\section{Model}\label{Model}

\begin{figure*}
\includegraphics[width=0.75\linewidth]{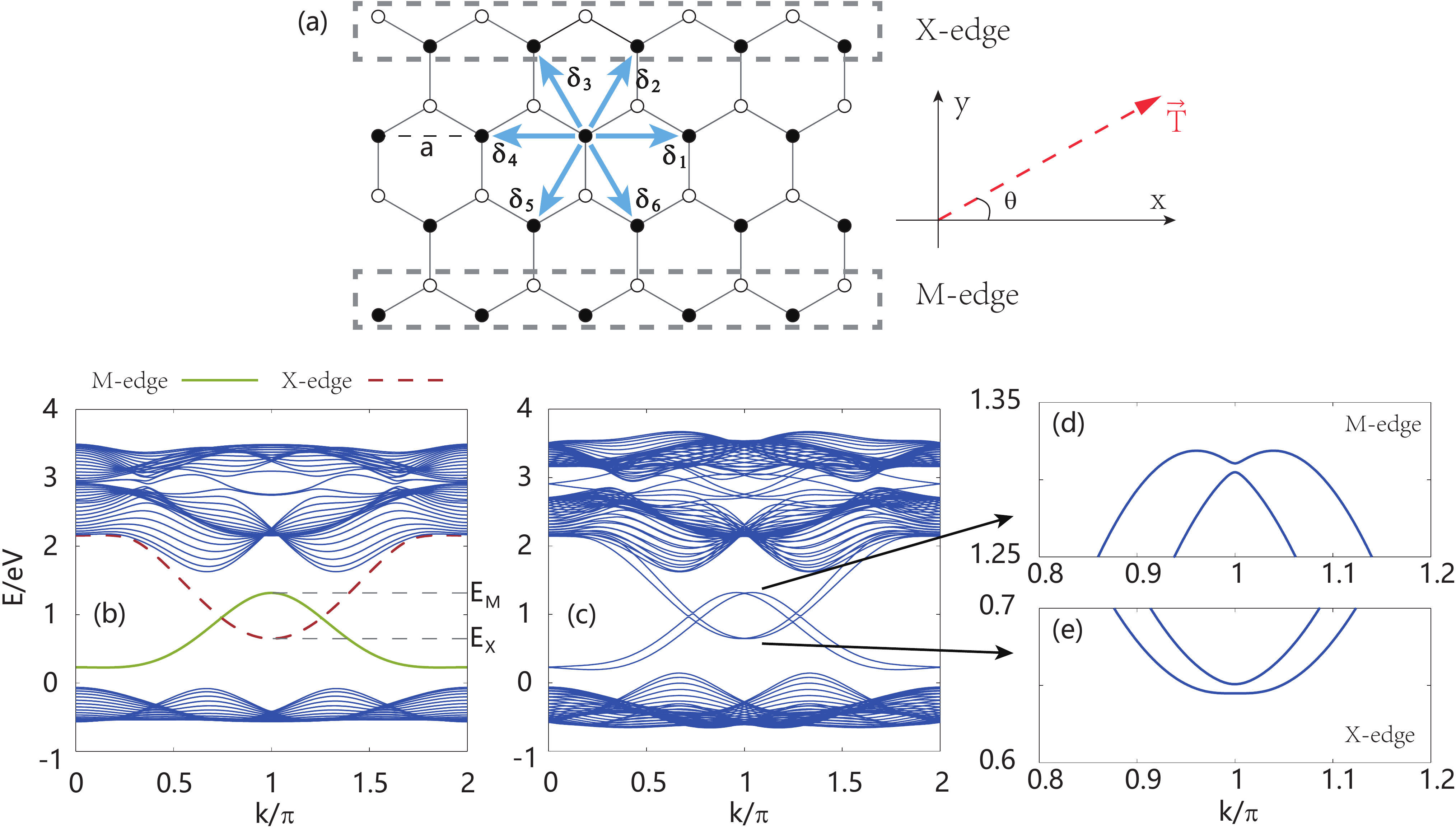}
\caption{Lattice structure and zigzag ribbon spectrum. (a) The zigzag ribbon structure of TMDs, with the zigzag direction chosen as the $x$ direction. Full circles are for the M-atoms and empty circles are for the X-atoms. The blue arrows show the nearest neighbor hoppings between M-atoms. The TMD ribbon is stretched or compressed along a prescribed direction, and $\theta$ is the direction of applied tension $\bm T$. (b) Band structure of a $\mathrm{MoS_2}$ zigzag ribbon with $N_y=20$ in the absence of SOC, magnetic field, superconductivity and strain. The green solid line and the red dashed line are for M-edge and X-edge, and $E_M$ and $E_X$ are the M-edge maximum energy and the X-edge minimum energy at $ka=\pi$, respectively. (c)-(e) Band structure with SOC $\lambda=460\,\mathrm{meV}$, Zeeman field $V_z=3\,\mathrm{meV}$, and zero superconductivity $\Delta = 0$. The number of lattice sites in $y$ direction is $N_y=20$. (d) and (e) show the magnetic gap on M-edge and X-edge respectively.}\label{fig1}
\end{figure*}

The system we consider is a monolayer $\mathrm{MX_2}$ zigzag ribbon deposited on top of an s-wave superconductor \cite{TMD_spin_SC}. The model has a honeycomb structure as showed in Fig.\ref{fig1}(a). Here we consider a three orbital model where only the M atom $d_{z^2}$, $d_{xy}$ and $d_{x^2-y^2}$ orbitals are taken into account. It has been shown that Bloch states of these materials around the gap edges are mostly contributed by these three orbitals \cite{3band}. The Hamiltonian can be written as,
\begin{eqnarray}
H=H_{0}+H_{SC}+H_{Z}-\mu N, \label{eq:Hamilt}
\end{eqnarray}
where $H_{0}$ is the 3-orbital tight binding Hamiltonian described below,  $H_{SC}$ is the superconducting paring term induced by proximity effect, $H_{Z}$ is the Zeeman splitting created by an in-plane magnetic field, $\mu$ is the chemical potential, and $N$ is the total electron  number operator. 

The first term $H_{0}$ is the 3-orbital tight-binding Hamiltonian of TMDs with a intrinsic SOC \cite{3band},  which for the 1D edge of the zigzag ribbon has an equivalent effect to the Rashba-type SOC in conventional 1D semiconducting nanowires \cite{TMD_spin_SC}. The Hamiltonian reads,
\begin{eqnarray}
H_{0} & = & \sum_{i,\bm{\delta}}\sum_{\gamma,\gamma',s}c_{i,\gamma,s}^{\dagger}t_{\bm{\delta},\gamma,\gamma'}c_{i+\bm{\delta},\gamma',s}+ \nonumber \\
 & + & \sum_{i,\gamma,s}c_{i,\gamma,s}^{\dagger}\epsilon_{\gamma}c_{i,\gamma',s}+i\lambda\sum_{i,s}\sum_{\underset{\gamma\neq\gamma'\neq d_{z^{2}}}{\gamma,\gamma'}}c_{i,\gamma,s}^{\dagger}\sigma_{z}^{ss}c_{i,\gamma',s}\,, \label{eq:H0}
\end{eqnarray}
where $c_{i,\gamma,s}^{\dagger}$ creates an electron at lattice site $i$ and orbital $\gamma$ with spin $s$, $\bm \delta $ are the six vectors connecting nearest neighbor M-atoms [see Fig.~\ref{fig1}(a)], $t_{\bm{\delta},\gamma,\gamma'}$ are hopping integrals, $\epsilon_\gamma$ are on-site energies, $\lambda$ is the SOC parameter, and $\sigma$ are the Pauli matrices acting on spin space. The parameters we choose are from first-principle calculation with generalized-gradient approximation for $\mathrm{MoS_2}$ in Ref.~\cite{3band}, with $\epsilon_{d_z²} \equiv \epsilon_1 =1.046$ and $\epsilon_{d_{xy}}=\epsilon_{d_{x²-y²}} \equiv \epsilon_2 =2.104$ in eV, and hopping amplitudes as given in Table.~\ref{tab1}.

The second term in Eq.~\eqref{eq:Hamilt} is for the s-wave induced superconductivity, and is given by,
\begin{eqnarray}
H_{SC}=\sum_{i,\gamma}\Delta c^{\dagger}_{i,\gamma,\uparrow} c^{\dagger}_{i,\gamma,\downarrow}+h.c.\,,
\end{eqnarray}
while the third term stands for an in-plane magnetic field, as a perpendicular one would induce vortices and break the translational symmetry. Without loss of generality, we choose a magnetic field in $x$ direction, and the Zeeman term is then given by,
\begin{eqnarray}
H_{Z}=V_z\sum_{i,\gamma,s,s'}c^{\dagger}_{i,\gamma,s}\sigma_x^{ss'}c_{i,\gamma,s'}\,.
\end{eqnarray}

\begin{table}
\begin{tabular}{|c|c|c|c|c|c|c|}
\hline
\hline
$\gamma$ - $\gamma'$
 & $\bm{\delta_1}$ & $\bm{\delta_2}$ & $\bm{\delta_3}$ & $\bm{\delta_4}$ &$\bm{\delta_5}$ & $\bm{\delta_6}$\\
\hline
$d_{z^2}$-$d_{z^2}$ & $t_{0}$ & $t_{0}$ & $t_{0}$ & $t_{0}$ & $t_{0}$ & $t_{0}$ \\
\hline
$d_{xy}$-$d_{xy}$ & $t_{11}$ & $\frac{t_{11}+3t_{22}}{4}$ & $\frac{t_{11}+3t_{22}}{4}$ & $t_{11}$ & $\frac{t_{11}+3t_{22}}{4}$ & $\frac{t_{11}+3t_{22}}{4}$ \\
\hline
$d_{x^2-y^2}$-$d_{x^2-y^2}$ & $t_{22}$ & $\frac{3t_{11}+t_{22}}{4}$ & $\frac{3t_{11}+t_{22}}{4}$ & $t_{22}$ & $\frac{3t_{11}+t_{22}}{4}$ & $\frac{3t_{11}+t_{22}}{4}$ \\
\hline
$d_{z^2}$-$d_{xy}$ & $t_{1}$ & $\frac{t_{1}+\sqrt{3}t_{2}}{2}$ & $-\frac{t_{1}+\sqrt{3}t_{2}}{2}$ & $-t_{1}$ & $-\frac{t_{1}-\sqrt{3}t_{2}}{2}$ & $\frac{t_{1}-\sqrt{3}t_{2}}{2}$ \\
\hline
$d_{xy}$-$d_{x^2-y^2}$ & $t_{2}$ &
$\begin{array}{c}\frac{\sqrt{3}(t_{11}-t_{22})}{4} \\ -t_{12}\end{array}$ & $\begin{array}{c}\frac{\sqrt{3}(t_{22}-t_{11})}{4} \\ +t_{12}\end{array}$ & $t_{2}$ & $\begin{array}{c}\frac{\sqrt{3}(t_{11}-t_{22})}{4} \\ +t_{12}\end{array}$ & $\begin{array}{c}\frac{\sqrt{3}(t_{22}-t_{11})}{4} \\ -t_{12}\end{array}$ \\
\hline
$d_{x^2-y^2}$-$d_{z^2}$ & $t_{12}$ & $-\frac{t_{2}+\sqrt{3}t_{1}}{2}$ & $-\frac{t_{2}+\sqrt{3}t_{1}}{2}$ & $-t_{12}$ & $-\frac{t_{2}-\sqrt{3}t_{1}}{2}$ & $-\frac{t_{2}-\sqrt{3}t_{1}}{2}$\\
\hline
\end{tabular}
\caption{The hopping amplitudes $t_{\bm{\delta},\gamma,\gamma'}$ in real space. Different rows show the hoppings between different orbitals, and different columns are for different spatial hopping vector $\bm \delta$ [see Fig.\ref{fig1}(a)].  The parameters are from the first-principle calculation with generalized-gradient approximation for $\mathrm{MoS_2}$ in Ref.~\cite{3band}: $t_0=-0.184$, $t_1=0.401$, $t_2=0.507$, $t_{11}=0.218$, $t_{12}=0.338$ and $t_{22}=0.057$ in eV.}\label{tab1}
\end{table}

Last but not least, we would like to incorporate the effect of an in-plane strain to this model. Strain will deform the lattice and change the distance between two lattice sites, hence only the terms in the Hamiltonian that connect different sites are affected. We model in-plane deformations by an in-plane deformation field $\mathbf{u}=(u_x,u_y)$. After deformation, the lattice point at $\mathbf{R}$ goes to a different position $\tilde{\mathbf{R}}$ such that $\tilde{\mathbf{R}} = \mathbf{R}+\mathbf{u}$. Accordingly, the non-deformed lattice distance between the  point at $\mathbf{R}$ and the point at $\mathbf{R}+\bm{\delta}$ may change after the deformation sets in. Under realistic conditions we expect $u_x , u_y \ll a $, where $a = |\bm \delta |$ is the lattice spacing. At linear order, the strain modified hopping may be written as,
\begin{eqnarray}
t_{\bm{\delta}}\rightarrow t_{\bm{\delta}}-t_{\bm{\delta}}\beta (\hat{\bm{\delta}}\cdot\bm{\nabla})\mathbf{u}\cdot\hat{\bm{\delta}},
\end{eqnarray}
where $\hat{\bm{\delta}}=\bm{\delta}/|\bm{\delta}|$ is a unit vector in the direction of $\bm \delta$, and the electron-phonon coupling parameter $\beta\approx -\partial \mathrm{log} t/\partial \mathrm{log} a \sim 3$   \cite{Roldan13,RRC+15}, similar to the case of graphene. In our model, we consider the six nearest neighbor hopping as showed in Fig.~\ref{fig1}(a), and they become,
\begin{eqnarray}
t\rightarrow\begin{cases}
t-t\beta u_{xx}, &\bm{\delta}=(\pm a,0);\cr t-t\beta (u_1+u_2), &\bm{\delta}=(\pm \frac{a}{2},\pm\frac{\sqrt{3}a}{2});
\cr t-t\beta (u_1-u_2), &\bm{\delta}=(\pm \frac{a}{2},\mp\frac{\sqrt{3}a}{2}), \label{hoppings}
\end{cases}
\end{eqnarray}
with $u_1=\frac{1}{4}u_{xx}+\frac{3}{4}u_{yy}$ and $u_2=\frac{\sqrt{3}}{2}u_{xy}$, and the strain tensor $u_{ab}=(\partial_a u_b+\partial_b u_a)/2$.


\section{Majorana Zero Modes and a topological invariant} \label{Berry}
We consider a zigzag ribbon  with open boundary conditions in $y$ direction, and set $k \equiv k_x$ in the following discussion. The ribbon has two nonequivalent edges, called X-edge and M-edge for obvious reasons, as shown in Fig.~\ref{fig1}(a).  Before adding strain, the spinless Hamiltonian has two edge states localized at the X-edge and M-edge respectively, as shown in Fig.~\ref{fig1}(b), which are the candidates to generate 1D topological Majorana states. For a spin-1/2 system, by adding the intrinsic SOC $\lambda$ and a in-plane magnetic field $V_z$, each of these edge bands will separate into two edge bands with opposite helicities, and a magnetic gap around the X-edge minimum energy $E_X$ or the M-edge maximum energy $E_M$ opens at $ka=\pi$, as depicted in Figs.~\ref{fig1}(c)-\ref{fig1}(e).

Following Ref.~\cite{TMD_spin_SC}, we mainly focus on the X-edge because of two reasons: First, according to the first principle calculations of Ref.~\cite{3band}, there are two spinless edge states of M-atoms at the M-edge, but the tight-binding model only gives one of them, while there is only one spinless  edge state from M-atoms along the X-edge in both tight-binding model and first principle calculations; Secondly, the M-edge is not as stable as the X-edge, as it can be greatly affected by edge passivations~\cite{XM-edge1,XM-edge2,XM-edge3,XM-edge4}.

\begin{figure}
\includegraphics[width=0.95\linewidth]{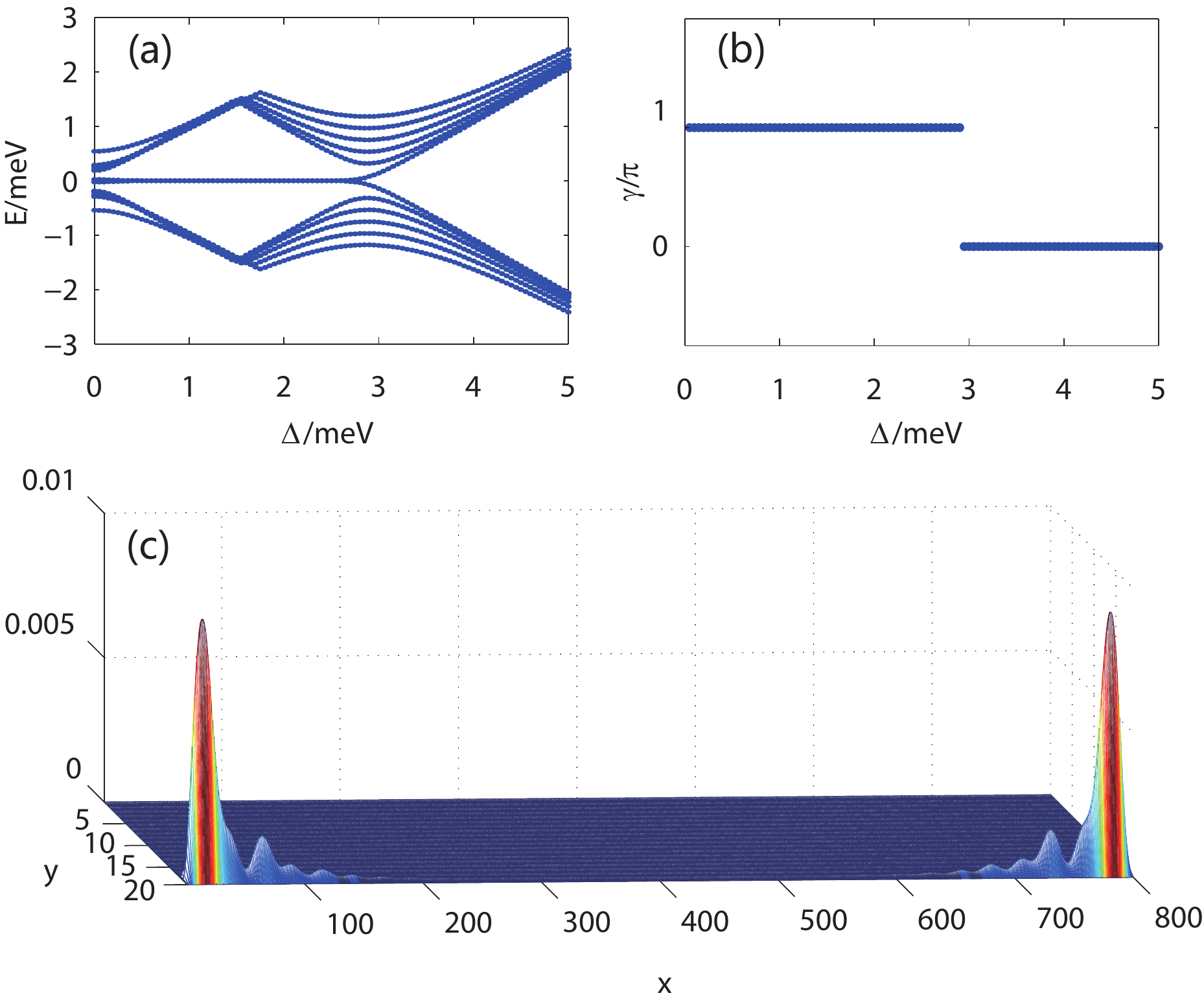}
\caption{Majorana zero modes and topological invariant. (a) Low energy spectrum versus $\Delta$. Only 12 modes closest to zero energy are plotted. (b) The Berry phase versus $\Delta$. Other parameters for (a) and (b) are $V_z=3\,\mathrm{meV}$  $\lambda=460\,\mathrm{meV}$, and the chemical potential $\mu=E_X\approx0.6479\mathrm{eV}$, which is between the gap in the X-edge. (c) The real-space distribution of the zero mode for $\Delta=1\,\mathrm{meV}$. The  lattice dimensions used in (a) and (c)  are $N_x=800$ and $N_y=20$.}\label{fig2}
\end{figure}

A nonzero superconducting order parameter $\Delta \neq 0$ makes the system nontrivial for $\Delta$ below some critical value, $\Delta < \Delta_c$. If the chemical potential is inside one of the gaps shown in Fig.~\ref{fig1}(d) and \ref{fig1}(e), then a pair of Majorana zero modes is induced on the corresponding edge. Here we choose the chemical potential to be inside the X-edge gap. The X-edge minimum energy $E_X$ satisfies the relation,
\begin{eqnarray}
(\epsilon_1-2t_0-E_X)(\epsilon_2-2t_{22}-E_X)-4t_2^2=0,\label{EX}
\end{eqnarray}
and the magnetic gap then develops from $E_X-V_z$ to $E_X+V_z$, as shown in the Appendix~\ref{appendix}. In this work, in order to guarantee that the Fermi level is in the gap, we choose $\mu=E_X$. The emergence of the zero-energy modes is shown in Fig.~\ref{fig2}(a), and Fig.~\ref{fig2}(c) shows the real-space distribution of the Majorana zero mode wave function. If we increase $\Delta$, the system will go through a topological phase transition when $\Delta > \Delta_c \approx V_z$, and becomes a trivial superconductor without Majorana zero modes, as shown in Fig.~\ref{fig2}(a).

According to the symmetry classification of topological systems this model belongs to the BdG class~D \cite{TopologicalClass}, as the Zeeman term breaks the time-reversal symmetry. The topology of this class in 1D can be characterized by a $Z_2$ number, which is associated with the Berry phase~\cite{D1,D2,D3}. In our system, we calculate the Berry phase $\gamma$ for the lower half bands to discriminate between topologically trivial and nontrivial phases. 

In order to calculate the Berry phase we consider $N$ discrete points within the 1D Brillouin zone,  $k \in [ 0, 2\pi[$, with  momentum $k$ taking values  $k_1,k_2,...,k_N$. Let us for the moment assume the system has a band well isolated from the remaining ones through finite gaps above and below. The band's Berry phase $\gamma$ may then be obtained by defining the link variable $U(k_l)=\varphi^{*}(k_l)\varphi(k_{l+1})$ and summing over $k_l$,
\begin{eqnarray}
\gamma= -i\sum_l\log{U(k_l)}\,, \label{eq:Berry}
\end{eqnarray}
where  $\varphi(k_l)$ are  the eigenstates of the Fourier transformed Hamiltonian $\mathcal{H}_k$ along the longitudinal direction of the ribbon, $\mathcal{H}_k \varphi(k_l) = E_k \varphi(k_l)$. If $n$ such bands are filled, the system's Berry phase is obtained by adding the Berry phases of the individual bands. This can be generalized to the case where the $n$ filled bands below some energy gap cross at degenerate points. Then we,  for the Berry phase of the lowest $n$ bands, we have $U(k_l)=\det{\mathbf{U}(k_l)}$ with
\begin{eqnarray}
\mathbf{U}_{ij}(k_l)=\varphi_i^{*}(k_l)\varphi_j(k_{l+1}),~1\leq i,j\leq n,
\end{eqnarray}
where $\varphi_i(k_l)$ is the eigenstate at momentum  $k_l$ associated to the $i$th band. 

Numerical results for the Berry phase of the present system are shown in Fig.~\ref{fig2}(b). It is clearly seen that  a finite Berry phase $\gamma = \pi $ in Fig.~\ref{fig2}(b) correlates with the presence of zero energy Majorana modes in Fig.~\ref{fig2}(a) for $\Delta < \Delta_c$. The transition to the trivial phase is signaled in Fig.~\ref{fig2}(a) by the absence of zero energy Majorana modes, and in Fig.~\ref{fig2}(b) by a zero Berry phase $\gamma = 0$. Note also that the present result fully agrees with the \emph{bulk - boundary} correspondence, as in Fig.~\ref{fig2}(a) we used a finite size system bounded in the $x$ direction ($N_x = 800$ and $N y = 20$), while in Fig.~\ref{fig2}(b) periodic boundary conditions along $x$ were used (keeping $N y = 20$).

\section{Result and discussion} \label{Results}

\subsection{The effect of strain}

Built in strain of the order $\lesssim 0.05\%$ seems to be unavoidable in current experiments \cite{Kumar15}. Even though such small deformations are spacially modulated, likely randomly, in order to set bounds on the effect of strain we use the simplest model with uniform strain. Despite simplicity, this approximation highly improves as the characteristic length scale of strain modulations increases. Moreover, it is shown below that strain can be used as a tuning parameter, and in this case, for experimental reasons, one of the easiest implementations is that of uniform, uniaxial strain. Another advantage of uniform strain is that it does not break the translational symmetry along the ribbon's longitudinal direction, and therefore the Berry phase in Eq.~\eqref{eq:Berry} is still well defined. 

For  uniform planar tension $\bm T$ [see Fig.~\ref{fig1}(a)], the strain tensor $u$ can be written in terms of the tensile strain $\varepsilon$ (relative deformation along the direction of $\bm  T$) as follows \cite{strain_graphene},
\begin{eqnarray}
u=\varepsilon\left(
\begin{array}{cc}
\cos^2{\theta}-\nu\sin^2{\theta} &(1+\nu)\cos{\theta}\sin{\theta}\\
(1+\nu)\cos{\theta}\sin{\theta} &\sin^2{\theta}-\nu\cos^2{\theta}
\end{array}\right),\label{tension}
\end{eqnarray}
where  $\theta$ is the angle between the direction of tension $\bm T$  and $x$ direction for the reference frame shown in Fig.\ref{fig1}(a), and $\nu$ is the Poisson's ratio for the material; $\nu=0.25$ for $\mathrm{MoS_2}$~\cite{PoissonRatio}. As before, we align $x$ with the zigzag direction and $y$ with the armchair direction. The strain tensor in Eq.~\eqref{tension} is used in Eq.~\eqref{hoppings} to determine the strain modified hoppings.

\begin{figure}
\includegraphics[width=0.95\linewidth]{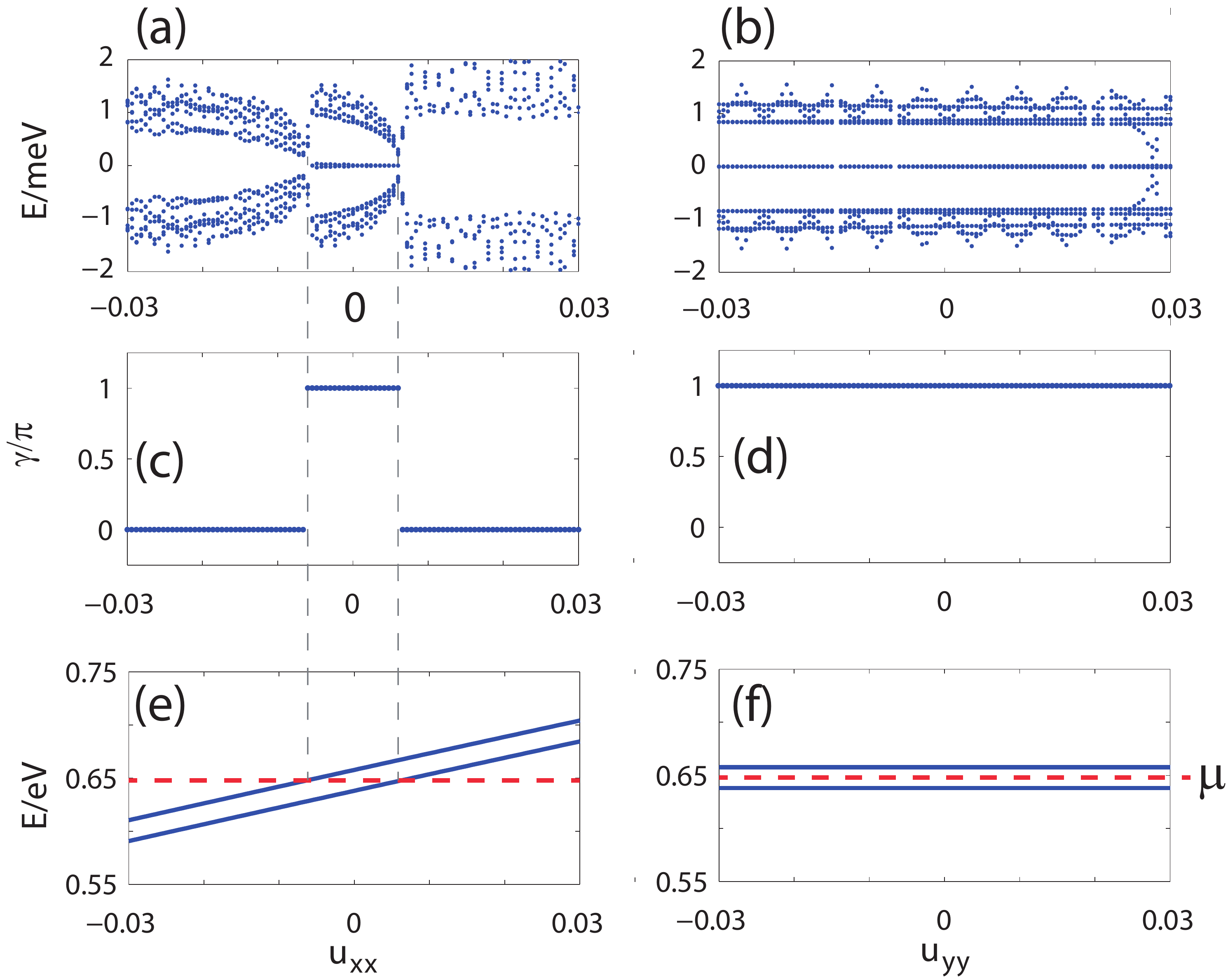}
\caption{The top row shows low energy spectrum as a function of $u_{xx}$ (a) and $u_{yy}$ (b) for open boundary conditions with $N_x=800$. Only the first 12 modes closest to zero are shown. Other parameters used are $\lambda=460\,\mathrm{meV}$, $V_z=10\,\mathrm{meV}$, $N_y=20$, and $\Delta=1\,\mathrm{meV}$. The middle row shows the Berry phase as a function of $u_{xx}$ (c) and $u_{yy}$ (d), respectively for the system in (a) and (b) with periodic boundary conditions; The bottom row presents  the magnetic gap at $ka=\pi$ for the X-edge without superconductivity as a function of $u_{xx}$ (e) and $u_{yy}$ (f). The gap is the region between the two blue solid lines. The red dashed line shows the chemical potential $\mu=E_X$, with $E_X$ the X-edge minimum energy without strain.}\label{fig3}
\end{figure}

Let us first consider the simplest case where we stretch or compress  only along $x$ or $y$ axis, and ignore the contraction extension in the transverse direction, i.e. $\theta=0$ or $\pi/2$ and the Poisson's ratio $\nu=0$. We find that a small deformation in $x$ direction ($u_{xx} \lesssim 0.01$) causes a topological phase transition to the trivial case, while deformations along $y$ direction do not affect the topological properties. The results are shown in Fig.~\ref{fig3}(a)-(d). Figures~\ref{fig3}(a) and \ref{fig3}(b) in the upper row show the spectrum for the finite system as a function of strain in the $x$ direction $u_{xx}$ and $y$ direction $u_{yy}$, respectively. Zero energy states, signaling the presence of Majorana modes, can be seen for deformations along $x$  only for $|u_{xx}| \lesssim 0.01$, while for deformations along $y$ the zero modes are present all over the region of probed strains. Results for the Berry phase are shown in Fig.~\ref{fig3}(c) and \ref{fig3}(d) in the middle row, respectively for $u_{xx} \neq 0$ and $u_{yy} \neq 0$. The Berry phase calculation completely agrees with the boundary analysis. In particular, the topological phase transition where the Berry phase goes from $\gamma=\pi$ to $\gamma = 0$, is clearly seen as a function of  $u_{xx}$ in Fig.~\ref{fig3}(c). Note that the results of the present section have been obtained for an unrealistically high Zeeman field, $V_z =10\, \mathrm{meV}$, just to make clear qualitatively how strain affects the system. Quantitative results are presented in Sec.~\ref{sec:robust}.

To understand these results, we go back to the case without superconductivity, $\Delta =0$, and see how strain changes the band structure. In Fig.~\ref{fig3}(e) it is shown that a nonzero $u_{xx}$ shifts the energy of the magnetic gap at X-edge, and the system becomes topologically trivial when the chemical potential $\mu$ falls outside the gap. In contrast, the value of $u_{yy}$ does not change the magnetic gap, as seen in Fig.~\ref{fig3}(f), and thus $u_{yy}$ cannot induce a topological phase transition \cite{M-edge_gap}.

\begin{figure}
\includegraphics[width=0.8\linewidth]{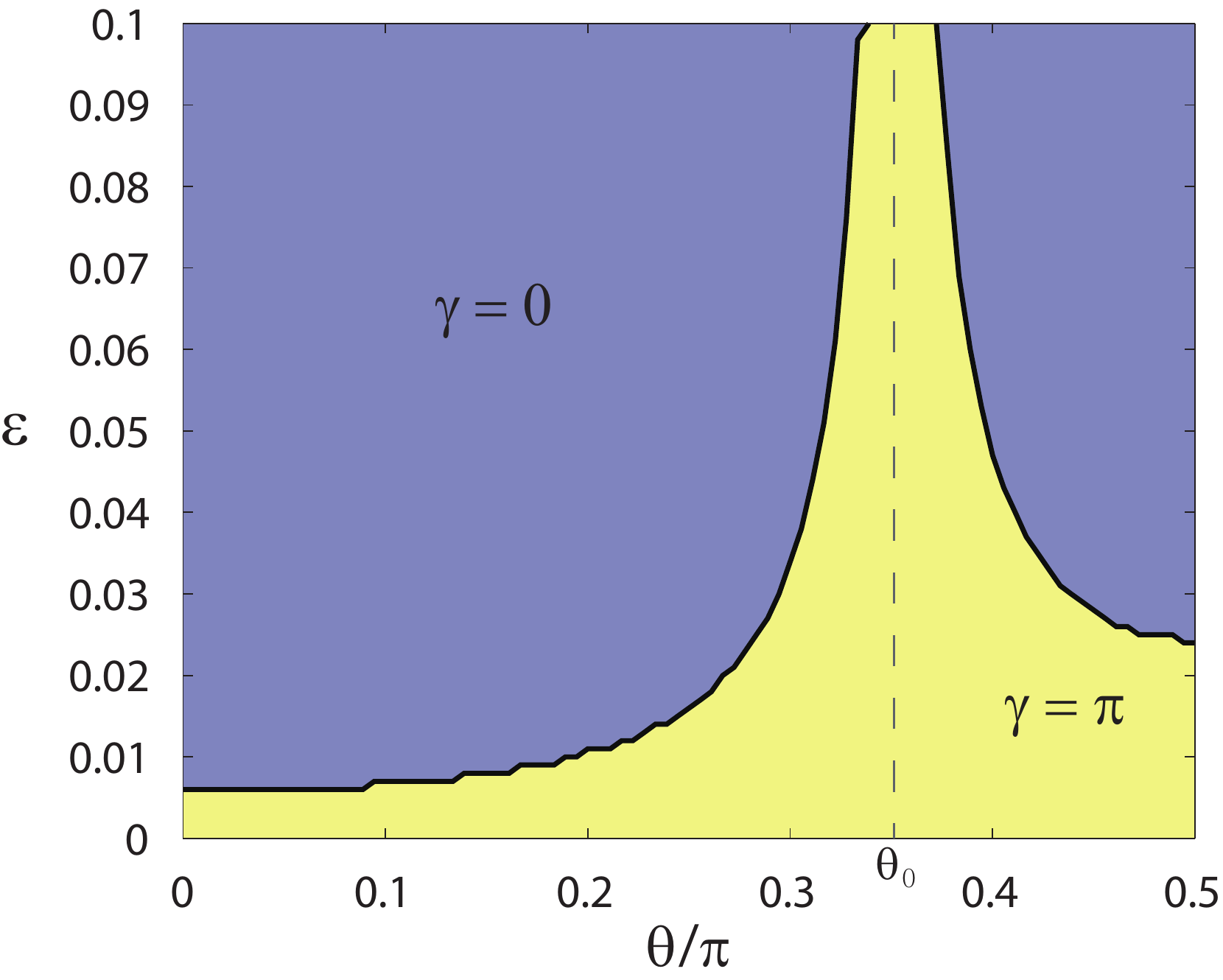}
\caption{Berry phase versus $\varepsilon$ and $\theta$ for $\mathrm{MoS_2}$, with the Poisson's ratio $\nu=0.25$, $\Delta=1\,\mathrm{meV}$, $V_z=10\,\mathrm{meV}$ and $\lambda=460\,\mathrm{meV}$. }\label{fig4}
\end{figure}

For a real material, the strain along a prescribed direction also causes deformation in the transverse direction, which is reflected by a nonzero Poisson's ratio $\nu$. In Fig.~\ref{fig4} the Berry phase is shown in the plane of tensile strain $\varepsilon$ and strain direction  $\theta$. The topologically nontrivial region increases from $\theta = 0$ to $\theta = \theta_0$,  narrowing down after this  critical $\theta_0$. The critical point is given by $u_{xx}=0$, which, from Eq.~\eqref{tension}, yields $\tan^2{\theta_0}=1/\nu$. Quantitative effects of strain are discussed in the next section for realistic values of parameters, in particular a realistic Zeeman field $V_z$.

\subsection{Robustness of the topological phase}\label{sec:robust}

\begin{figure}
\includegraphics[width=0.95\linewidth]{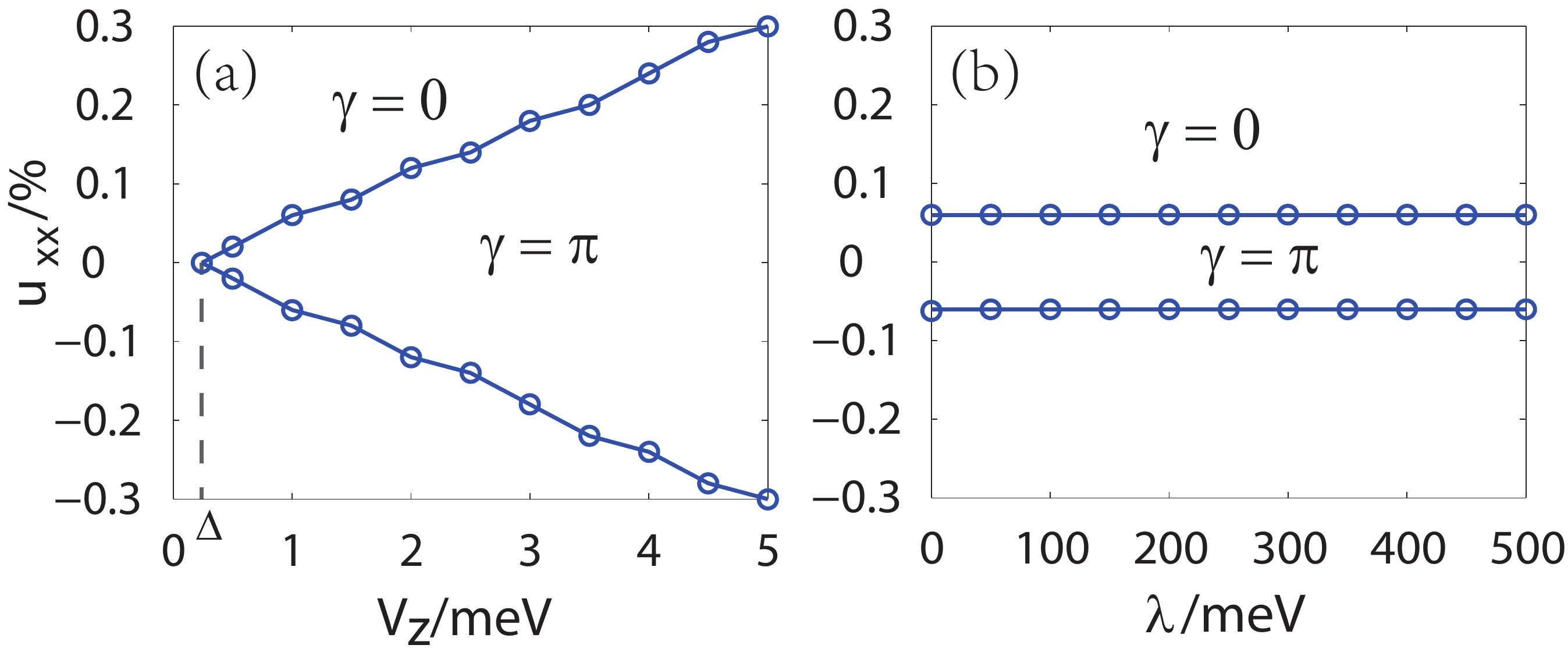}
\caption{Critical strain strength as a function of Zeeman field  $V_z$ (a) and SOC parameter $\lambda$ (b). The system has a Berry phase $\gamma=\pi$ in the topological side of the phase diagram, which changes to $\gamma = 0$  in the trivial phase. The jump $\Delta \gamma = \pi $ defines the phase boundaries, which are given by the two blue lines in each figure. In~(a) we fixed $\lambda=460\,\mathrm{meV}$, and in~(b) we fixed $V_z=1\,\mathrm{meV}$. The superconducting order parameter is $\Delta=0.25\,\mathrm{meV}$ for both figures.}\label{fig5}
\end{figure}

Here we discuss quantitative effects of strain for the present system under realistic conditions. We only consider the deformation along $x$ direction, as it has the strongest effect.

As shown in Sec.~\ref{Berry}, the magnetic gap for the X-edge at $ka=\pi$ exactly equals $V_z$. We therefore expect a larger $V_z$, which induces a wider magnetic gap, to enhance the topologically nontrivial region. This is precisely shown in  Fig.~\ref{fig5}(a), where the phase diagram in the plane $V_z$ versus $u_{xx}$ is presented. The topological phase occurs around $u_{xx}=0$, and the critical strain roughly increases linearly with $V_z$. Note that the value of $V_z$  needs to be larger than the induced superconducting order parameter, which for a typical $s$-wave superconductor such as NbTiN is of the order of $\Delta\approx 0.25\, \mathrm{meV}$  \cite{SC_experiment}. The value of $\Delta$ is also indicated in Fig.~\ref{fig5}(a). The SOC parameter $\lambda$, on the contrary,  does not affect the critical strain strength, as shown in Fig.~\ref{fig5}(b). Based on these quantitative  results, we may establish three different regimes regarding the effect of strain on the Majorana zero modes of TMDs at zigzag edges:

(i)~For small Zeeman fields, typically $V_z < 1 \, \mathrm{meV}$, the effect of strain strongly limits the presence of Majorana modes. Even small residual strain of order $\varepsilon \lesssim 0.05 \%$ \cite{Kumar15} is sufficient to completely destroy the topological phase. 
For realistic cases, built in strain is expected to be randomly distributed not only in strength but also in direction, with regions where even biaxial strain may dominate. Based on the result shown in  Fig.~\ref{fig4}, we anticipate that a randomly distributed strain could give rise to a situation of phase separation, where topological  regions located around strain minima are surrounded by non-topological ones pinned at strain maxima. A percolating driven transition could then be induced by increasing the applied magnetic field, or other conveniently chosen external parameter. These findings put important constraints on the realization of Majorana physics at the zigzag edges of TMDs \cite{TMD_spin_SC}. The following two points may be used to circumvent such limitations.

(ii)~For moderate Zeeman fields, typically $V_z \gtrsim 1 \, \mathrm{meV}$, the critical strain is above the current built in strain limit of order $\varepsilon \lesssim 0.05 \%$ \cite{Kumar15}, and well below the maximum strain $\varepsilon \gtrsim 10 \%$ \cite{elasticity}. Strain may then be used as a tunable parameter to turn the Majorana mode phase on and off. Note, however, that a Zeeman field $V_z$ of $1\sim2$~meV, which would overcome the effect of a strain $\varepsilon \sim 0.1\%$, requires a magnetic field of several Tesla for a typical $g-$factor~2. One should then carefully choose the experimental setup, in order to not destroy the superconductivity of the nearby superconducting material used in the proximity effect. 

(iii)~Finally, the effect of uniform strain one can be made irrelevant by tuning the chemical potential. The magnetic gap extends between $E_X-V_z$ and $E_X+V_z$, with strain only affecting the X-edge minimum energy $E_X$. This means that strain does not narrow down the topological nontrivial region. A shift in the chemical potential $\mu$, for instance through an electric field effect, is then enough to guarantee  the existence of Majorana zero modes. Therefore, we can also control the Majorana modes and avoid the effect of uniform strain by tuning the chemical potential to a proper value. In Fig.~\ref{fig3}(c)-(f) we can see that the system holds Majorana zero modes and has a Berry phase $\gamma=\pi$ as long as the chemical potential is inside the gap. The width of the gap is given by $2V_z$, and the center of the gap depends on the value of $u_{xx}$. For non-uniform strain this strategy is obviously much less effective, and one must then resort to minimize strain by conveniently chose the substrate.

Let us point out that the value $V_z\sim1 \, \mathrm{meV}$ needed to crossover from regime (i) to regime (ii) comes from the measured built in strain in current experiments, in particular Ref.~\cite{Kumar15} where residual strain $\lesssim 0.05\%$ was clearly demonstrated. For smaller built in strain we expect the crossover to occur for a reduced Zeeman field. Given the linear relation between the critical strain and Zeeman [Fig.~\ref{fig5}(a)], the $V_z$ value needed for  crossover may change considerably. Based on results for graphene \cite{flatness, OCK+12}, we anticipate that a carefully chosen substrate could substantially reduce the critical strain (a facor of~10 seems feasible, given the height to length ration of characteristic deviations from flatness). One substrate worth trying is BN, which has been shown to greatly improve the electronic properties of two-dimensional materials~\cite{substrate}.

\section{Conclusions}\label{Conclusions}
We have studied the effect of strain in the 1D topological phase realized at the zigzag edges of transition-metal dichalcogenides. In-plain deformations turn out to strongly affect the topologically nontrivial region in parameter space. Depending on the applied magnetic field and on whether the chemical potential is kept fixed or tuned, we have identified three distinct regimes: (i) For fixed chemical potential  and not so high magnetic field (Zeeman field $V_z < 1\, \mathrm{meV}$), built in strain $\lesssim 0.05\%$ \cite{Kumar15} is enough to completely wash out the topological phase. This puts severe limits on the realization of 1D Majorana physics in this system \cite{TMD_spin_SC}. A conveniently chosen substrate wich keeps buil-in strain to a minimum is mandatory in this case. (ii) For fixed chemical potential and high magnetic field (Zeeman field $V_z \gtrsim 1\, \mathrm{meV}$), strain can be used to tune the topological phase by turning it on for small strain and off by increasing strain within experimentally accessible values. (iii) For fixed, uniform strain, the effects of strain can be avoided by adjusting the system's chemical potential through electric field effect. 

Finally, we have further calculated the Berry phase as the topological invariant of this system, and demonstrated the correspondence between the Berry phase and the existence of the Majorana modes. 

Strain engineering of Majorana modes is a unique feature of two-dimensional materials since they sustain high deformations within the elastic regime \cite{Zhenhua}. Transition metal dichalcogenides appear as good candidates to realize this phenomenon. 

\begin{acknowledgements}
Partial support from FCT-Portugal through Grant No.~UID/CTM/04540/2013 is acknowledged.
\end{acknowledgements}

\begin{appendix}

\section{X-edge analytics}\label{appendix}
In this appendix we calculate the X-edge minimum energy and the magnetic gap analytically. The momentum $k_x$ is a good quantum number in this model, hence we take the eigenstate of the Hamiltonian in $k_x-y$ space as

\begin{eqnarray}
\Psi_{k_x}=\sum_{y,s} (\phi_{k_x,y}^{a,s}\hat{a}^{\dagger}_{k_x,y}+\phi_{k_x,y}^{b,s}\hat{b}^{\dagger}_{k_x,y}+\phi_{k_x,y}^{c,s}\hat{c}^{\dagger}_{k_x,y})|0\rangle,
\end{eqnarray}
with $|0\rangle$ the vacuum state and $s$ the index of spin, $\hat{a}^{\dagger}$, $\hat{b}^{\dagger}$ and $\hat{c}^{\dagger}$ are the creation operator of orbital $d_{z^2}$, $d_{xy}$ and $d_{x^2-y^2}$, respectively. Here we focus on the X-edge minimum point of $k_x a=\pi$, and write $\phi_{\pi,y}$ as $\phi_{y}$ for convenience. By requiring $H_0\Psi_{k_x}=E_{k_x}\Psi_{k_x}$, we have the eigen-equations
\begin{widetext}
\begin{eqnarray}
&&(\epsilon_1-2t_0)\phi_{y}^{a,s}+V_z\phi_{y}^{a,s'}-2t_2\phi_{y}^{c,s}+i(t_1+\sqrt{3}t_2)\phi_{y+1}^{b,s}+i(t_1-\sqrt{3}t_2)\phi_{y-1}^{b,s}=E\phi_{y}^{a,s},\nonumber\\
&&(\epsilon_2-2t_{22})\phi_{y}^{c,s}+V_z\phi_{y}^{c,s'}-2t_2\phi_{y}^{a,s}-i[\frac{\sqrt{3}}{2}(t_{22}-t_{11})-2t_{12}]\phi_{y+1}^{b,s}-i[-\frac{\sqrt{3}}{2}(t_{22}-t_{11})-2t_{12}]\phi_{y-1}^{b,s}=E\phi_{y}^{c,s},\nonumber\\
&&(\epsilon_2-2t_{11})\phi_{y}^{b,s}+V_z\phi_{y}^{b,s'}-i(t_1-\sqrt{3}t_2)\phi_{y+1}^{a,s}-i(t_1+\sqrt{3}t_2)\phi_{y-1}^{a,s}\nonumber\\
&&~~~~~~~~~~~~~~~~~~~~+i[-\frac{\sqrt{3}}{2}(t_{22}-t_{11})-2t_{12}]\phi_{y+1}^{c,s}+[\frac{\sqrt{3}}{2}(t_{22}-t_{11})-2t_{12}]\phi_{y-1}^{c,s}=E\phi_{y}^{b,s},\label{eigen}
\end{eqnarray}
\end{widetext}
Here we only consider the original three band model of $\mathrm{MX_2}$ with a Zeeman field $V_z$. When $V_z=0$, the two components of spin decouple, and we find that the X-edge eigenstate is given by $\phi_{y}^{b}=0$ for any $y$ and
\begin{eqnarray}
\frac{\phi_{y}^{a}}{\phi_{y}^{c}}&=&\frac{2t_2}{\epsilon_1-2t_0-E_X}=\frac{\epsilon_2-2t_{22}-E_X}{2t_2},\label{Xedge}\\
\frac{\phi_{y-1}^{a}}{\phi_{y+1}^{a}}&=&-\frac{M_1}{M_2},
\end{eqnarray}
with $M_1=[-\frac{\sqrt{3}}{2}(t_{22}-t_{11})-2t_{12}]\frac{\epsilon_1-2t_0-E}{2t_2}-(t_1-\sqrt{3}t_2)$ and $M_2=[\frac{\sqrt{3}}{2}(t_{22}-t_{11})-2t_{12}]\frac{\epsilon_1-2t_0-E}{2t_2}-(t_1+\sqrt{3}t_2)$, and the eigenenergy $E_X$ satisfies Eq.~\eqref{EX}.

For the parameters of $\mathrm{MoS}_2$, we have the X-edge minimum energy $E_X=0.6479\mathrm{eV}$ (the other solution of $E$ is mixed with bulk states). Note decaying ratio $\phi_{y-1}^{a}/\phi_{y+1}^{a}\thickapprox0.0379$, which shows that the eigenstate is well localized at the X-edge.

For a finite Zeeman coupling, the ratio between $\phi_{y}^{a}$ and $\phi_{y}^{c}$ becomes,
\begin{eqnarray}
\frac{\phi_{y,s}^{a}}{\phi_{y,s}^{c}}&=&\frac{2t_2(\epsilon_1+\epsilon_2-2t_0-2t_{22}-2E_X)}{(\epsilon_1-2t_0-E_X)^2+4t_2^2-V_z^2}\nonumber\\
&=&\frac{(\epsilon_2-2t_{22}-E_X)^2+4t_2^2-V_z^2}{2t_2(\epsilon_1+\epsilon_2-2t_0-2t_{22}-E_X)}.\label{Xedge2}
\end{eqnarray}
By solving this equation, we can show that the two spin full states running at the X-edge have eigenenergies  $E_X\pm V_z$, which determine the magnetic gap.

Finally, the strain strength will only change the hopping amplitudes in the Hamiltonian, so that the hoppings appearing  in Eqs.~\eqref{Xedge} and~\eqref{Xedge2} become $t\rightarrow t(1-\beta u_{xx})$. This modification will only affect the value of $E_X$, while the magnetic gap is still determined by $E_X\pm V_z$.

\end{appendix}



\begin{thebibliography}{99}
\bibitem{TQC}
C. Nayak, S. H. Simon, A. Stern, M. Freedman, and S. Das Sarma, Rev. Mod. Phys. \textbf{80}, 1083 (2008).

\bibitem{Lutchyn}
R. M. Lutchyn, J. D. Sau, and S. Das Sarma, Phys. Rev. Lett. \textbf{105}, 077001 (2010).
\bibitem{Oreg}
Y. Oreg, G. Refael, and F. von Oppen, Phys. Rev. Lett. \textbf{105}, 177002 (2010).

\bibitem{SC_experiment}
V. Mourik, K. Zuo, S. M. Frolov, S. R. Plissard, E. P. A. M. Bakkers, and L. P. Kouwenhoven, Science \textbf{336}, 1003 (2012).
\bibitem{SC_experiment2}
M. T. Deng, C. L. Yu, G. Y. Huang, M. Larsson, P. Caroff, and H. Q. Xu, Nano Lett. \textbf{12}, 6414 (2012).
\bibitem{SC_experiment3}
A. Das, Y. Ronen, Y. Most, Y. Oreg, M. Heiblum, and H. Shtrikman, Nat. Phys. \textbf{8}, 887 (2012).
\bibitem{SC_experiment4}
L. P. Rokhinson, X. Liu, and J. K. Furdyna, Nat. Phys. \textbf{8}, 795 (2012).

\bibitem{XXH14}
Xiaodong Xu, Wang Yao, Di Xiao, and Tony F. Heinz, Nat. Phys. \textbf{10}, 343 (2014).

\bibitem{TMD}
B. Radisavljevic, A. Radenovic, J. Brivio, V. Giacometti, and A. Kis, Nat. Nanotechnol. \textbf{6}, 147 (2011);
Q. H. Wang, K. Kalantar-Zadeh, A. Kiss, J. N. Coleman, and M. S. Strano, Nat. Nanotechnol. \textbf{7}, 699 (2012);
H. Nam, S. Wi, H. Rokni, M. Chen, G. Priessnitz, W. Lu, and X. Liang, ACS Nano \textbf{7}, 5870 (2013).

\bibitem{TMD_spin_SC}
R.-L. Chu, G.-B. Liu, W. Yao, X. Xu, Di Xiao, and C. Zhang, Phys. Rev. B \textbf{89}, 155317 (2014).

\bibitem{TMD_Majorana}
G. Xu, J. Wang, B. Yan and X.-L. Qi, Phys. Rev, B \textbf{90} 100505 (2014).

\bibitem{Potter}
A. C. Potter and P. A. Lee, Phys. Rev. B \textbf{85}, 094516 (2012).
\bibitem{Sau}
J. D. Sau, S. Tewari, and S. Das Sarma, Phys. Rev. B \textbf{85}, 064512 (2012).

\bibitem{Wang}
 L. Wang, A. Kutana, and B. I. Yakobson, Ann. Phys. \textbf{526}, L7 (2014).
\bibitem{Feng}
 J. Feng, X. Qian, C.-W. Huang, and J. Li, Nat. Photon. \textbf{6}, 866 (2012).

\bibitem{Roldan13}
A. Castellanos-Gomez, Rafael Rold\'an Emmanuele Cappelluti, Michele Buscema, 
Francisco Guinea, Herre S. J. van der Zant, and Gary A. Steele,
Nano Lett. \textbf{13}, 5361 (2013).

\bibitem{Kumar15}
S. Kumar, A. Kaczmarczyk, and B. D. Gerardot, Nano Lett. \textbf{15}, 7567 (2015).

\bibitem{Scalise}
 E. Scalise, M. Houssa, G. Pourtois, V. Afanas¡ev, and A. Stesmans, Nano Res. \textbf{5}, 43 (2012).
\bibitem{G-Asl}
 M. Ghorbani-Asl, S. Borini, A. Kuc, and T. Heine, Phys. Rev. B \textbf{87}, 235434 (2013).

\bibitem{Ochoa}
M. A. Cazalilla, H. Ochoa, and F. Guinea, Phys. Rev. Lett. \textbf{113}, 077201 (2014).

\bibitem{RRC+15}
H. Rostamani, R Rold\'an, E. Cappelluti, R. Asgari, F. Guinea, Phys. Rev. B \textbf{92}, 
195402 (2015).

\bibitem{COK+10}
Eduardo V. Castro, H. Ochoa, M. I. Katsnelson, R. V. Gorbachev, D. C. Elias, K. S. Novoselov, and A. K. Geim; F. Guinea,
Phys. Rev. Lett. \textbf{105}, 266601 (2010).

\bibitem{OCK+12}
H. Ochoa, Eduardo V. Castro, M. I. Katsnelson, F. Guinea,
Phys. E \textbf{44}, 963 (2012).

\bibitem{RYZ+13}
C. Rice, R. J. Young, R. Zan, U. Bangert, D. Wolverson, T. Georgiou, R. Jalil, and K. S. Novoselov, 
Phys. Rev. B \textbf{87}, 081307(R) (2013).

\bibitem{3band}
G.-B. Liu, W.-Y. Shan, Y. Yao, W. Yao and D. Xiao, Phys. Rev. B \textbf{88}, 085433 (2013).

\bibitem{XM-edge1}
Z. Wang, H. Li, Z. Liu, Z. Shi, J. Lu, K. Suenaga, S. Joung, T. Okazaki, Z. Gu, J. Zhou, Z. Gao, G. Li, S. Sanvito, E. Wang, and S. Iijima, J. Am. Chem. Soc. \textbf{132}, 13840 (2010);
\bibitem{XM-edge2}
Y. Li, Z. Zhou, S. Zhang, and Z. Chen, J. Am. Chem. Soc. \textbf{130}, 16739 (2008);
\bibitem{XM-edge3}
H. Pan and Y. Zhang, J. Mater. Chem. \textbf{22}, 7280 (2012);
\bibitem{XM-edge4}
E. Erdogan, I. H. Popov, A. N. Enyashin, and G. Seifert, Eur. Phys. J. B \textbf{85}, 33 (2012).

\bibitem{TopologicalClass}
A. P. Schnyder, S. Ryu, A. Furusaki and A. W. W. Ludwig, Phys. Rev. B, \textbf{78} 195125 (2008).
S. Ryu, A. P. Schnyder, A. Furusaki and A. W. W. Ludwig, New J. Phys., \textbf{12} 065010 (2010).

\bibitem{D1}
Y. Hatsugai, J. Phys. Soc. Jpn. \textbf{75}, 123601 (2006).
\bibitem{D2}
Budich J. C. and Ardonne E., Phys. Rev, B  \textbf{88}, 075419 (2013).
\bibitem{D3}
L. Li, C. Yang and S. Chen, arXiv:1512.07386

\bibitem{strain_graphene}
V. M. Pereira, A. H. Castro Neto, and N. M. R. Peres, Phys. Rev. B \textbf{80}, 045401 (2009).

\bibitem{PoissonRatio}
D. M. Guzman and A. Strachan, J. Appl. Phys. \textbf{115}, 243701 (2014).

\bibitem{M-edge_gap}
A nonzero $u_{yy}$ changes the position of the magnetic gap of M-edge. If the chemical potential is placed in this gap, the system can also support Majorana zero modes, but they are eliminated by a finite $u_{yy}$.

\bibitem{elasticity}
A. Castellanos-G\'omez, M. Poot, G. A. Steele, H. S. J. van der Zant, N. Agra\"it, and G. Rubio-Bollinger, 
Adv. Mater. \textbf{24}, 772 (2012).

\bibitem{flatness}
Chun Hung Lui, Li Liu, Kin Fai Mak, George W. Flynn, Tony F. Heinz, 
Nature \textbf{462}, 339 (2009).

\bibitem{substrate}
A. V. Kretinin, Y. Cao, J. S. Tu, G. L. Yu, R. Jalil, K. S. Novoselov, S. J. Haigh, A. Gholinia, A. Mishchenko, M. Lozada, T.  Georgiou, C. R. Woods, F. Withers, P. Blake, G. Eda, A. Wirsig, C. Hucho, K. Watanabe, T. Taniguchi, A. K. Geim, and R. V. Gorbachev, 
Nano Lett. \textbf{14}, 3270 (2014).

\bibitem{Zhenhua}
Zhen-Hua Wang, Eduardo V. Castro, and Hai-Qing Lin,
arXiv:1601.05326, unpublished.


\end{thebibliography}
\end{document}